\documentclass{nature}
\usepackage{graphicx, color}
\usepackage{xcolor}
\usepackage{amsfonts,amssymb,amsmath}
\usepackage{siunitx}
\usepackage{bm}
\usepackage{bbold}
\usepackage{braket}
\usepackage{mathtools}
\usepackage{dsfont}
\usepackage{pdfpages}

\usepackage[normalem]{ulem}

\newcommand*{\YA}{\textcolor{black}}

\title{Long-range crossed Andreev reflection in topological insulator nanowires proximitized by \YA{a superconductor}}

\author{Junya Feng$^{1}$, Henry F. Legg$^{2}$, Mahasweta Bagchi$^{1}$, Daniel Loss$^{2}$, Jelena Klinovaja$^{2}$ \& Yoichi~Ando$^1$}

\makeatletter
\let\saved@includegraphics\includegraphics
\AtBeginDocument{\let\includegraphics\saved@includegraphics}
\renewenvironment*{figure}{\@float{figure}}{\end@float}
\makeatother

\begin{document}

\maketitle

\begin{affiliations}
 \item Physics Institute II, University of Cologne, Z\"ulpicher Str. 77, 50937 K\"oln, Germany
 \item Department of Physics, University of Basel, Klingelbergstrasse 82, CH-4056 Basel, Switzerland
\end{affiliations}

\begin{abstract}
Crossed Andreev reflection (CAR) is a nonlocal transport phenomenon that creates/detects Cooper-pair correlations between distant places. It is also the basis of Cooper-pair splitting to generate remote entanglement. Although CAR has been extensively studied in semiconductors proximity-coupled to a superconductor, it has been very difficult to observe it in a topological insulator (TI). Here we report the first observation of CAR in a proximitized TI nanowire (TINW). We performed local and nonlocal conductance spectroscopy on mesoscopic TINW devices in which superconducting (Nb) and metallic (Pt/Au) contacts are made on a bulk-insulating TINW. The local conductance detected a hard gap, accompanied by the appearance of Andreev bound states that can reach zero-bias, while a negative nonlocal conductance was occasionally observed upon sweeping the chemical potential, giving evidence for CAR. Surprisingly, the CAR signal was detected even over 1.5 $\mu$m, which implies that pair correlations extend over a length scale much longer than the expected superconducting coherence length of either Nb or the proximitised TINW. Such a long-range CAR effect is possibly due to an intricate role of disorder in proximitized nanowires. Also, our 0.9-$\mu$m device presented a decent Cooper-pair splitting efficiency of up to 0.5.
\end{abstract}


Topological superconductors\cite{Sato2017} have been a subject of significant interest due to their relevance to non-Abelian Majorana zero-modes that are the key to topological quantum computation\cite{Nayak2008}. 
Proximitized Rashba nanowires (NWs)\cite{Oreg2010, Lutchyn2010} have been leading the experimental efforts in this direction\cite{Prada2020, Flensberg2021, Aghaee2023}, but the realization of Majorana zero-modes remains controversial\cite{Valentini2021, Dassarma2023, Hess2023}, mainly due to disorder and the ambiguity of signatures observed in experimental realizations. Crossed Andreev reflection (CAR), which is a nonlocal transport phenomenon reflecting superconducting (SC) correlations between two normal-metal leads connected to a superconductor, has been suggested as a probe of bulk topological superconductivity in Rashba NWs due to its expected independence from local physics\cite{Rosdahl2018, Menard2020, Danon2020, Puglia2021, Pan2021, Hess2021, Pikulin2021, Wang2022, Poschl2022, Maiani2022, Banerjee2023PRL}, but without clear success so far\cite{Hess2023}. 
CAR is also used for creating Andreev molecule \cite{Su2017} and simulating topological superconductivity in the minimal Kitaev chain realized in semiconductor double quantum dots\cite{Dvir2023}, besides being used for 
Cooper-pair splitting\cite{Recher2001, Hofstetter2009, Sato2010, Deacon2015} in a similar setting.

Topological superconductivity would be more naturally realized in topological insulators (TI)\cite{Fu2008} than in Rashba NWs, because a large Zeeman energy is not necessary in TIs for realizing a topological gap. Motivated by this prospect, the SC proximity effect in a TI has been studied in various structures\cite{Zhang2011, Maier2012, Williams2012, Wang2012, Finck2014, Hart2014, Rosenbach2021, Bai2022, Uday2024}. Proximitized TINWs are particularly interesting in this respect, because theory predicts the appearance of Majorana zero-modes in the presence of moderate parallel magnetic fields\cite{Cook2011, Cook2012, deJuan2019, Legg2021, Heffels2023}. However, experimental studies of the SC proximity effect in TINWs have been mostly performed on Josephson junctions and the obtained information has been limited\cite{Rosenbach2021, Bai2022, Fisher2022, Roessler2023, Kim2017, Bhattacharyya2018}, let alone any signature of CAR. To better understand the SC proximity effect in TINWs, we turn to a combination of local and nonlocal conductance spectroscopies \cite{Gramich2017} and tried to detect the CAR signature \cite{Fuchs2021}, as has been done for proximitized Rashba NWs \cite{Rosdahl2018, Menard2020, Danon2020, Puglia2021, Pan2021, Hess2021, Pikulin2021, Wang2022, Poschl2022, Maiani2022, Banerjee2023PRL, Aghaee2023}.
In the local conductance, we found a hard gap can occur, signalling a robust proximity-induced superconductivity, along with the appearance of Andreev bound states (ABSs) that arise in the unproximitised normal sections at the end of the nanowire and can reach zero-energy even in zero magnetic field. Our measurement of the nonlocal conductance found that our SC-TINW hybrid device works as a decent Cooper-pair splitter and, surprisingly, that CAR can take place over length scales of as long as 1.5 $\mu$m.
This implies that an SC correlation that is necessary for the CAR process can be established throughout the TINW for a length scale much longer than the expected SC coherence length $\xi_{\rm SC}$, which is estimated to be shorter than $\sim$ 110 nm. This long-range CAR effect goes beyond simple descriptions of nonlocal conductance in mesoscopic superconductors and points to an intricate role of disorder in proximitized NWs; namely, overlapping ABSs created by disorder can lead to an ``Andreev band'' that extends beyond $\xi_{\rm SC}$ and support long-range CAR\cite{Hess2023}. As such, our result is an interesting example against the common belief that nonlocal transport phenomena should be independent of local physics and provide a simple measure of bulk SC properties.

\begin{flushleft}
{\bf Experimental set-up}
\end{flushleft}
\vspace{-8mm}

To perform the local and nonlocal conductance spectroscopy \cite{Gramich2017} of a proximitized TINW, we prepared devices as shown in Fig. 1a: We first fabricated a TINW by dry-etching\cite{Roessler2023} an exfoliated flake ($\sim$15-nm thickness) of the bulk-insulating TI material BiSbTeSe$_2$ (BSTS)\cite{Ren2011} to have a width of 100--200 nm, and then superconducting Nb electrodes and metallic Pt/Au electrodes were deposited onto the TINW with a gap of about 500 nm between these electrodes. The TINW shown in Fig. 1a was used for most of the measurements reported in this paper and had the cross-section of about 140 $\times$ 11 nm$^2$. Our finding was essentially reproducible in a total of 5 similar nonlocal devices. The local and nonlocal conductance spectroscopy was performed by using the Pt/Au electrodes made on either side of a grounded Nb electrode, as depicted in Fig. 1a. The local and nonlocal differential conductances $G_{ji} \equiv dI_j/dV_i$ ($i, j$ = L, R) define a conductance matrix: $V_i$ is the bias voltage applied to the left (L) or right (R) of the TINW via the Pt/Au electrode, and the resulting $I_j$ is measured as a function of $V_i$. In the actual experimentt, we superimposed a small-amplitude ac voltage $v^i_{ac}$ of frequency $f_i$ to the dc bias $V_i$ to modulate the chemical potential $\mu$ on the $i$-side, and the resulting ac component of the current $i^j_{ac}$ at the $j$ electrode was lock-in detected at the frequency $f_i$ to calculate the differential conductance $G_{ji} \equiv dI_j/dV_i = i^j_{ac}/v^i_{ac}$. The sign of $I_j$ was defined to be positive when the current flows into (out from) the middle electrode for the local (nonlocal) conductance. The details of the measurement circuit is shown in the supplement. 
The device had a global back gate. While we expect that the chemical potential, $\mu$, is essentially pinned in the sections covered by metals (Nb and Pt/Au), the $\mu$ in the section between them (bare-TINW) is tunable by the back-gate voltage $V_{\rm g}$. 

At 8 K where the Nb electrode ($T_c \approx$ 6.5 K) is in the normal state, the presence of the subbands characteristic of TINWs can be inferred from the Aharonov-Bohm (AB)-like oscillations in $G_{\rm LL}$ in parallel magnetic fields $B_{\parallel}$ \cite{Peng2010, Zhang2010, Xiu2011}, and these oscillations occasionally present $\pi$-phase shifts as $V_{\rm g}$ is changed \cite{Cho2015, Jauregui2016}, leading to a ``checkerboard'' pattern in the $B_{\parallel}$ vs $V_{\rm g}$ plane \cite{Kim2020, Roessler2023} as shown in Fig. 1b. The periodicity of the AB-like oscillations corresponds to the flux quantum $\phi_0 = h/e$, and the observed $\sim$2.7 T is reasonable for the cross-section of our TINW. 

\begin{flushleft}
{\bf Electron confinement and Andreev bound states}
\end{flushleft}
\vspace{-8mm}

The $V_{\rm g}$-dependence of $G_{\rm LL}$ at 8 K in zero magnetic field shown in Fig. 1c indicates that the Dirac point is reached at $V_{\rm g} \approx -5$ V. At the base temperature of 17 mK, however, we observed that $G_{\rm LL}$ measured at zero bias oscillates strongly as a function of  $V_{\rm g}$. To understand this oscillations, we took the differential conductance spectra with a small $V_{\rm g}$ step (Figs. 1d), which clarified that the strong $G_{\rm LL}$ oscillations is due to repeated opening and closing of a gap, which take place with a spacing $\Delta V_{\rm g}$ of only $\sim$0.01 V. The electron-hole symmetry in $G_{\rm LL}$ as a function of $V_{\rm L}$ strongly suggests that the origin of the gap is superconductivity; however, the nearly-regular gap closing as a function of $V_{\rm g}$ is unusual, suggesting that some mesoscopic effect is at work.

The strong sensitivity of the gap to $V_{\rm g}$ suggests that this gap exists in the section of the TINW not covered by Nb nor Pt/Au; namely, we are likely seeing the proximity-induced states in the bare-TINW section, which is $\sim$500 nm in length. When the TINW is brought into proximity with Nb, in addition to inducing superconductivity, metallization effects drastically change the TINW properties in the region under the Nb contact, for instance resulting in a change of Fermi-level and Fermi-velocity~\cite{Legg2022}. We can also expect similar effects, albeit likely weaker, for the TINW under the Pt/Au contact. The result is that a potential well forms in the bare-TINW section between the  Pt/Au contacts and Nb, i.e. in the N region of the N'NS' structure between normal lead and superconducting lead (see Fig.~\ref{fig:Figure2}). Similar to Rashba NWs \cite{Reeg2018, Escribano2018, Cayao2021}, this potential well results in the formation of ABSs in this bare TINW region. The energy of each ABS depends on the relative strengths of normal reflection and Andreev reflection at the Nb interface [see Fig.~\ref{fig:Figure2}(a) and (b), respectively]. Dominant Andreev reflection results in a bound state energy close to the SC gap edge, whereas dominant normal reflection results in a bound state energy within the induced SC gap. The origin of the strong $V_{\rm g}$ dependence of $G_{\rm LL}$ now becomes clear: As the chemical potential is adjusted, the ratio of normal and Andreev reflection is altered. Indeed, for an ABS resulting from a single subband with Fermi-momentum $k_F$, Fabry-Perot-like interference can result in a large normal reflection when the length $L$ of the N-section satisfy the resonance condition $2k_{\rm F} L = (N_{\rm FP}+\frac{1}{2})\pi$, with $N_{\rm FP}$ an integer \cite{Reeg2018, Cayao2021}. 
To be more concrete, in Fig.~\ref{fig:Figure2}(c) we show a simulation of the energy of a single bound state in an NS region (see supplement for details). As expected, as the chemical potential is altered the energy of the ABS oscillates within the superconducting gap depending on the ratio of Andreev and normal reflection. We note that the energy never reaches precisely zero, which at zero-magnetic field is possible only for perfect normal reflection  \cite{Reeg2018}. However, in Fig.~\ref{fig:Figure2}(e) we show the outcome of an example simulation (see supplement) of the local conductance $G_{\rm LL}$ without a well defined tunnel barrier to the normal lead (Pt/Au contact). Due to the weak tunnel barrier, a strong coupling of the ABS to the normal lead results in a significant broadening of the local conductance signature such that it can reach zero-bias, as observed in our experiment.
\YA{Note that the separability of the transverse physics from longitudinal physics in TINW\cite{Legg2021} allows us to use a one-dimensional model for the above simulations; the transverse modes are quantized and indexed by angular momentum, which affects the longitudinal modes only indirectly\cite{Legg2021, Legg2022}.}

One can check the consistency of this scenario as follows: the Dirac point is at $V_{\rm g} \approx -5$~V (Fig. 1c) and the $V_{\rm g}$-separation between resonances, $\Delta V_{\rm g}$, seen in Fig. 1d near $V_{\rm g} = -3$ V is $\sim$0.01~V. Thus, there should be roughly 200 resonances between $-5$ V and $-3$ V. This implies $N_{\rm FP} \approx$ 200 at $V_{\rm g} = -3$ V, giving $k_{\rm F} \approx$ 0.6 nm$^{-1}$ for $L$ = 500 nm. With the Fermi velocity $v_{\rm F} = 3 \times 10^5$ m/s for the Dirac cone in BSTS \cite{Arakane2012}, this $k_{\rm F}$  corresponds to $E_F \approx$ 0.1 eV, which is in the right ballpark. In addition, the estimate of $\sim$200 resonances for $E_F \approx$ 0.1 eV implies an energy interval of $\sim$0.5 meV, which is consistent with our observation that the checker-board pattern in $G_{\rm LL}(B_{\parallel}, V_{\rm g})$ gradually breaks up into more fine-structured patterns for $T \lesssim$ 4 K (see supplement).
In passing, from the temperature dependence of the AB-like conductance oscillations, we estimate the phase-coherence length $L_{\rm p}$ of 1.9 $\mu$m in our TINW (see supplement). This supports the existence of Fabry-Perot-like interference in the 500-nm-long bare-TINW.

\begin{flushleft}
{\bf Nonlocal conductance and crossed Andreev reflection}
\end{flushleft}
\vspace{-8mm}

While the local $G_{\rm LL}$ turns out to be mainly governed by the states in the N-section of our device, the nonlocal $G_{\rm RL}$ contains information on the S'-section. The spectra of $G_{\rm RL}(V_{\rm L})$ measured for the same $V_{\rm g}$ range are plotted in Fig. 1f (the bias was applied only to the left lead; namely, $V_{\rm R}$ = 0 in this measurement). One can see that at zero bias, $G_{\rm RL}$ vanishes when $V_{\rm g}$ is off-resonance, while a positive $G_{\rm RL}$ is observed at on-resonance. This positive $G_{\rm RL}$ is most likely due to the process called elastic co-tunneling (ECT), in which an electron tunnels between the N-sections on the left and right of S'; since an electron injected from the left comes out as an electron on the right, $G_{\rm RL}$ is positive in the case of ECT. A competing process is CAR, in which an electron injected into S' from the left forms a Cooper pair by taking an electron from the right of S', ejecting a hole into the right N-section; this hole causes a negative $G_{\rm RL}$ and the Cooper pair created in S' is drained to the ground (Fig. 3a). The CAR process requires a state that is extended throughout the S' part \cite{Rosdahl2018}, and therefore CAR gives unambiguous evidence for the SC correlation in S'. 
\YA{Our data in Fig. 1f show that CAR can become dominant at the $V_g$ positions marked by blue and green ticks, where $G_{\rm LL}$ (see Fig. 1e) suggests the existence of a sufficient number of states at $V_L> 0$ on the left of S'; note that, being a nonlocal process, the occurrence of CAR depends also on the ABSs on the right of S', which is reflected in $G_{\rm RR}$. The orange tick marks a $V_g$ position at off-resonance (Fig. 1d), where the absence of states at low bias (Fig. 1e) leads to $G_{\rm RL} \approx$ 0 (Fig. 1g).}
This is the first observation of CAR in a proximitized TINW. Note that the length of S' is 900 nm here, while the SC coherence length $\xi_{\rm Nb}$ in our Nb is about 30 nm \cite{Uday2024}, meaning that CAR must be taking place through the proximitized TINW, not through Nb.
 
The competition between ECT and CAR can be manipulated by applying a finite bias not only to the left but also to the right (i.e. making $V_{\rm R} \neq 0$) \cite{Bordin2023}. For example, when $V_{\rm L} = V_{\rm R} < 0$, the process to split a Cooper pair in S' into one electron on each side (Fig. 3b) is enhanced. This Cooper-pair splitting is microscopically the same process as CAR and it gives rise to negative $G_{\rm RL}$ and $G_{\rm LR}$. The inverse of the Cooper-pair splitting, i.e. electrons from either side creating a Cooper pair in S' (Fig. 3c), is promoted when  $V_{\rm L} = V_{\rm R} > 0$, and this process also yields negative $G_{\rm RL}$ and $G_{\rm LR}$. On the other hand, with $V_{\rm L} = -V_{\rm R}$, the process of an electron on the high-bias side to simply tunnel through S' to the low-bias side is enhanced; this is microscopically the ECT process and yields positive $G_{\rm RL}$ and $G_{\rm LR}$. 
\YA{In our device, we measured all four differential conductances, $G_{\rm LL}$, $G_{\rm RL}$, $G_{\rm LR}$ and $G_{\rm RR}$, in the $V_{\rm L}$ vs $V_{\rm R}$ plane at a fixed global $V_{\rm g}$. The results shown in Figs. 3d-3g, obtained with a $V_{\rm g}$ where
$G_{\rm LL}$ is at off-resonance and $G_{\rm RR}$ is relatively small, confirm the expected controllability of the competition, although the behavior in Fig. 3e is less ideal (see supplement).} 
(We note that we used different frequencies $f_L$ and $f_R$ for the left and right bias modulations, respectively, and measured $i^R_{ac}$ at the frequency $f_L$, such that $G_{\rm RL} =  i^R_{ac}/v^L_{ac}$ is made to reflect the bias drive on the left; the same is true for $G_{\rm LR}$.)
Taking advantage of this fact, we measured the full conductance matrix by setting $V_{\rm L} = V_{\rm R}$ to enhance CAR and the results are shown in Fig. 4, which presents a $V_{\rm g}$ range near the Dirac point. 
\YA{One can see that negative $G_{\rm LR}$ and $G_{\rm RL}$ signaling the CAR process appear frequently upon sweeping $V_{\rm g}$.} 
The range of $V_{\rm g}$ where CAR occurs is broadly consistent between $G_{\rm RL}$ and $G_{\rm LR}$, confirming that this process reflects the SC coherence throughout the S'-section of the TINW.

\begin{flushleft}
{\bf Effects of parallel magnetic fields}
\end{flushleft}
\vspace{-8mm}

A parallel magnetic field generates a magnetic flux $\varphi$ in the TINW and modulates the electronic structure in a nontrivial way. In particular, the subband gap closes and a spin-non-degenerate band appears at $\varphi =\left(n+\frac{1}{2}\right)h/e$. This change is expected to have the following implications for the proximity effect: on one hand, it has been suggested that the odd number of bands is ideal for inducing topological superconductivity~\cite{Cook2011}; on the other hand, the flux of $\left(n+\frac{1}{2}\right)h/e$ shifts the angular momentum of the subbands in such a way that the Cooper pairing becomes forbidden (called angular-momentum mismatch)~\cite{deJuan2019}, at least in a system with rotational symmetry. To see the effects of $\varphi = \frac{1}{2}(h/e)$ in our device, we measured the full conductance matrix at this flux with a fixed bias of $V_{\rm L} = V_{\rm R} = -0.05$ mV for the same ${V}_{\rm g}$ range as in Fig. 4, and the result is shown in Fig. 5. 
One can see that there is no drastic change in the data and the key features, the periodic gap opening/closing and the existence of CAR, are still observed at $\varphi = \frac{1}{2}(h/e)$. When we look into details, the gap becomes soft and CAR is observed less frequently, which is consistent with a reduced, but nonzero, superconducting gap. The $\varphi$-dependence of the conductance matrix for a range of $V_{\rm g}$ at a fixed bias is also consistent with a reduced gap near $\varphi = \frac{1}{2}(h/e)$ (see supplement).

\begin{flushleft}
{\bf Absence of Majorana zero modes}
\end{flushleft}
\vspace{-8mm}

Our data for $\varphi = \frac{1}{2}(h/e)$ clearly show that the SC gap does not close at this flux, which implies that the angular-momentum mismatch does not completely prohibit the SC gap in our TINW. This result confirms the theoretical prediction \cite{deJuan2019, Legg2021, Legg2022, Heffels2023} that the inversion symmetry breaking caused by having the SC contact only on the top surface and/or gating from one side is sufficient for mitigating the mismatch problem, which is a good news for future experiments.
When Cooper paring is present in the TINW at $\varphi = \frac{1}{2}(h/e)$, one would expect the appearance of a topological SC state accompanied by Majorana bound states (MBSs) at any chemical potential\cite{Cook2011}. However, we do not observe a stable zero-bias conductance peak (ZBCP) that would signal the MBSs.

The absence of a robust ZBCP in our experiment does {\it not} necessarily mean that a topological SC state is not induced in our TINWs. 
The dominance of the ABS in the bare-TINW section likely results in a weak coupling to the leads of any MBS state formed at the end of the SC section. Furthermore, the presence of CAR and ECT indicates that extended ABSs are present in the SC section, and these would hybridise the two MBSs expected to form at the ends of this section. In other words, although topological superconductivity might be realized, it is not surprising that ZBCPs do not appear in our setup. This issue should be settled by future experiments in which the normal sections at the end of the TINW are reduced in length with well defined tunnel barriers and a longer TINW is used to reduce the hybridisation of the potential MBSs.
\YA{Since there is a large room for improvements in hybrid TINW devices that are more tolerant of disorder than Rashba-nanowire devices\cite{Legg2021}, it remains interesting to pursue MBSs in TINWs.}

\begin{flushleft}
{\bf Devices with different lengths of the S'-section}
\end{flushleft}
\vspace{-8mm}

To understand how the local and nonlocal conductances change when the length $L_{\rm S'}$ of the S'-section is varied, we show the results from devices with $L_{\rm S'}$ = 1.5 and 0.5 $\mu$m. Figure 6 shows $G_{LL}$ and $G_{RL}$ measured by applying only $V_{\rm L}$ to the device (i.e. $V_{\rm R}$ = 0). Firstly, the $V_{\rm g}$-distance between the resonances in both devices is nearly the same as that in the $L_{\rm S'}$ = 900 nm device. This supports our interpretation that the gap closing is due to the Fabry-Perot-like resonance in the N-section, whose length was always around 500 nm. Secondly, the CAR signal was observed in the $G_{RL}$ data of even the $L_{\rm S'}$ = 1.5 $\mu$m device, which means that the SC correlation in the S'-section extends to the length scale of at least 1.5 $\mu$m. 

To examine the implications of our observation, we estimate the SC coherence length $\xi_{\rm SC}$ in the TINW. 
\YA{If we take the observed gap of $\sim$0.2 meV seen in the orange curve in Fig. 1e as the lower bound of the induced SC pairing potential $\Delta$ (which can be larger if the orange curve is dominated by in-gap states in N or S'),} 
the upper bound of the clean-limit BCS coherence length $\xi_{\rm BCS} = \hbar v_F/(\pi \Delta)$ is estimated to be $\sim$0.3 $\mu$m, which is already shorter than 1.5 $\mu$m. Since our BSTS has a mean free path $\ell_{\rm mfp}\approx$ 40 nm \cite{Taskin2011}, a cautious estimate gives $\xi_{\rm SC} = \sqrt{\xi_{\rm BCS}\ell_{\rm mfp}}$ of at most $\sim$110 nm. Therefore, our 1.5-$\mu$m device shows that CAR can be observed for the length scale more than $\sim$10 times longer than $\xi_{\rm SC}$. 
At first sight such a very long-range CAR effect is surprising, since it suggests the existence of subgap states that have a spatial extent much longer than $\xi_{\rm SC}$. In this regard, it was recently shown theoretically that quasi-periodic disorder in a proximitized nanowire can create overlapping ABSs that form a band extending throughout the nanowire and result in a long-range nonlocal conductance that can have a dominant CAR contribution\cite{Hess2023}. 
\YA{Our observation points to the existence of such overlapping ABSs in the S' region, connecting the left and right N regions coherently for the CAR to take place.}
For $L_{\rm S'} \sim 10\, \xi_{\rm SC}$ observed in our TINWs, this would require the overlap of only a few ABSs. This scenario seems highly plausible in our setup, since the relatively high level of disorder\cite{Huang2021} in our compensated TINWs likely results in a considerable number of local ABSs when proximitized.


\begin{flushleft}
{\bf Discussions and conclusions}
\end{flushleft}
\vspace{-8mm}


The Coulomb blockade in a quantum dot can open a hard gap and hence one might think that it would be premature to conclude that the hard gap we observed is due to superconductivity. In this regard, the $G_{\rm LL}$ spectra shown in Fig. 1d show that the on-resonance states appear vertically in this plot and they do not disperse with $V_{\rm G}$, particularly in the region where the resonances appear regularly. This behavior is not consistent with the Coulomb blockade and supports the Fabry-Perot-like origin of the gap closing.

The very first observation of the CAR process in a TI platform was reported recently for a ferromagnetic TI realizing the quantum anomalous Hall effect, where the CAR happened across a narrow Nb electrode contacting the chiral edge state\cite{Uday2024}. Combined with the present result, it appears that a dominant CAR contribution is more easily achieved when the transport channel is one-dimensional.

It is useful to mention that the nonlocal conductance is sometimes advertised to be free from complications coming from local physics and allows us to probe the topological phase throughout the nanowire \cite{Microsoft2023}. In this regard, our data show that $G_{\rm RL}$ and $G_{\rm LR}$ are primarily governed by the local states formed in the N-sections, and the nonlocal physics plays a secondary role in determining them. Nevertheless, it is interesting to see that when the local states in the N-sections allow the CAR process, our device can work as a Cooper pair splitter. If we define the Cooper-pair-splitting efficiency $\eta_{\rm CPS} \equiv i_{ac}^j / i_{ac}^i$ ($i \neq j$) for the $v_{ac}^i$ drive with frequency $f_i$, we obtain $\eta_{\rm CPS}$ of up to 0.5 for our $L_{\rm S'}$ = 900 nm device. This is not as high as that obtained in dedicated Cooper-pair splitters (up to 0.9)\cite{Schindele2012}, but is still a decent efficienty \cite{Bordin2023, Dvir2023}.

In conclusion, we found that mesoscopic TINW devices in which both Pt/Au and Nb leads contact the TINW, metallization effects can result in a potential well that allows the formation of ABSs in the bare-TINW section between normal and superconducting leads. We demonstrated that Fabry-Perot-like interference result in an apparent gap closing that shows up when a resonance condition is met, while a hard SC gap at off-resonance signals a robust SC proximity effect. Most importantly, we observed negative nonlocal conductance across the Nb-covered section of the TINW, which is a signature of CAR and indicates that the SC correlation extends throughout the Nb-covered section. This CAR signal was observed even in a device with 1.5-$\mu$m-long Nb-covered section, which is more than 10 times longer than $\xi_{\rm SC}$. 
This surprisingly long-range CAR effect is possibly due to overlapping ABSs\cite{Hess2023} that are created by disorder and can .extend much beyond $\xi_{\rm SC}$. In addition, we found that it is possible to induce superconducting correlations, evinced by CAR, even for $\varphi = \frac{1}{2}(h/e)$ where topological superconductivity is expected. Overall, our data demonstrate that nonlocal transport in mesoscopic SC systems reflects a complex interplay of local and global properties, and understanding this interplay in the presence of disorder is vital for using nonlocal transport as a probe of exotic proximity-induced superconductivity.

\begin{flushleft}
{\bf Methods}
\vspace{-5mm}

{\bf Device fabrication:} Large BSTS flakes were exfoliated from a bulk single crystal and then transferred deterministically to a doped Si substrate with a 290-nm thick SiO$_2$ coating layer. We use electron-beam lithography (EBL) to define the nanowire pattern and Ar dry-etching to fabricate nanowires from the flakes. Nb leads and Pt/Au leads were fabricated separately with a standard EBL process and UHV sputter-deposition; 
\YA{a small overlay error between the two steps caused slight differences in the lengths of the N regions.} 
Before metallization, the nanowire surface was cleaned by an {\it ex-situ} O$_2$ etching and an {\it in-situ} Ar-plasma cleaning. \YA{Having the N regions of TINW between Nb and Pt/Au gives the tunability of $G_{\rm LL}$ and $G_{\rm RR}$, which is important for the observation of CAR. With non-tunable tunnel barriers, CAR is difficult to be observed.}

{\bf Measurements:} The devices were measured in an Oxford Instruments Trtion 300 dilution refrigerator equiped with a 6-1-1-T superconducting vector magnet. Standard low-frequency lock-in technique with dc bias was used to measure the differential conductance (see supplement for details). 

{\bf Theory:} Simulations were done using the Python package KWANT~\cite{KWANT} for a one-dimensional nanowire with a single subband and different chemical potential between S' or N' and N regions (see supplement for details). 
\YA{Note that the surface-state wavefunction of the TINW decays almost completely in $\sim$3 nm,\cite{Legg2022} so it form the shape of a rectangular tube; the analysis in Ref. \citenum{Legg2022} suggests that its thickness is not strongly modified by the contact with Nb or gating.}

\end{flushleft}

\begin{addendum}

\item[Acknowledgments:] 
This project has received funding from the European Research Council (ERC) under the European Union’s Horizon 2020 research and innovation program (Grant Agreement No. 741121) and was also funded by the Deutsche Forschungsgemeinschaft (DFG, German Research Foundation) under Germany's Excellence Strategy - Cluster of Excellence Matter and Light for Quantum Computing (ML4Q) EXC 2004/1 - 390534769, as well as by the DFG under CRC 1238 - 277146847 (Subprojects A04 and B01). H.F.L. acknowledges support by the Georg H. Endress Foundation.

\item[Author contributions:] 
Y.A. concieved the project. J.F. perfomed device fabrication, measurements and data analysis with the help of Y.A. M.B. grew BSTS crystals. H.F.L., J.K. and D.L. performed theoretical analysis. Y.A., J.F. and H.F.L. wrote the manuscript with input from all authors.

\item[Competing Interests:] The authors declare no competing interests.

\item[Correspondence:] Correspondence should be addressed to Y.A. (ando@ph2.uni-koeln.de).

{\bf Data and code availability:} Raw data and codes used in the generation of main and supplementary figures are available in Zenodo with the identifier 10.5281/zenodo.12611922.

\end{addendum}

\clearpage
\begin{figure}[tb]
    \centering
    \includegraphics[width=\textwidth]{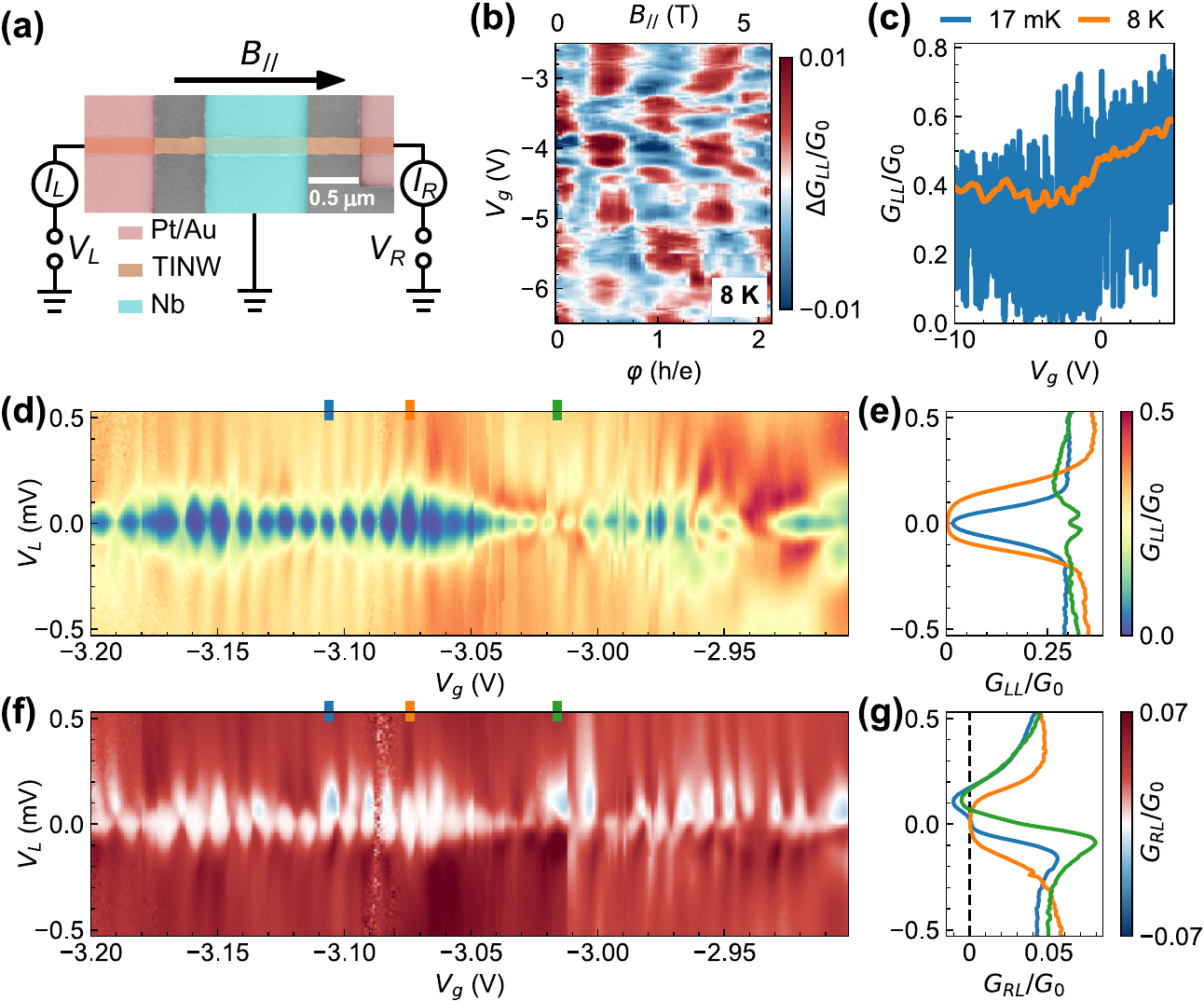}
	\caption{\linespread{1.05}\selectfont{}
	{\bf Local and nonlocal conductance of a TINW proximitized by Nb.}
	{\bf (a)} False-color scanning electron microscope (SEM) image of the device and a schematics the measurement setup. Scale bar: 0.5 $\mu$m.
	{\bf (b)} Checker-board pattern in \YA{zero-bias} $G_{\rm LL}$ in the $B_{\parallel}$ vs $V_{\rm g}$ plane 
	\YA{at 8 K, which is above the $T_c$ of the Nb electrode; here, we plot $\Delta G_{\rm LL}$ which is calculated by subtracting a smooth $B_{\parallel}$-dependent background (see supplement for details) from $G_{\rm LL}$ to enhance the visibility of the oscillations.} 
	The flux generated by $B_{\parallel}$ in the TINW is shown on the bottom axis in the unit of the flux quantum $h/e$, which corresponds to $B_{\parallel}$ = 2.7 T. $G_0 \equiv 2e^2/h$.
	{\bf (c)} $V_{\rm g}$-dependence of zero-bias $G_{\rm LL}$ at 8 K and 17 mK. In the 8-K data, one can see that  $G_{\rm LL}$ presents a broad mininum at around $V_{\rm g} = -5$ V, which signals the Dirac point.
	{\bf (d, f)} Plots of $G_{\rm LL}$ and $G_{\rm RL}$ measured as a function of $V_{\rm L}$ and $V_{\rm g}$ at 17 mK with $V_{\rm R}$ = 0. 
	{\bf (e, g)} Line cuts of the $G_{\rm LL}$ and $G_{\rm RL}$ spectra at three $V_{\rm g}$ values marked by coloured ticks ($-3.106$, $-3.074$ and $-3.016$ V) in panels (d) and (f). The orange spectrum of $G_{\rm LL}$ in (e) is an example of the hard gap. The green and blue spectra of $G_{\rm RL}$ in (g) go \YA{negative} at small positive bias, which is a signature of CAR;
	\YA{here, $V_{\rm L}$ is varied with $V_{\rm R}$ = 0.}
    }
    \label{fig:Figure1}
\end{figure}

\begin{figure}[b]
    \centering
    \includegraphics[width=1\textwidth]{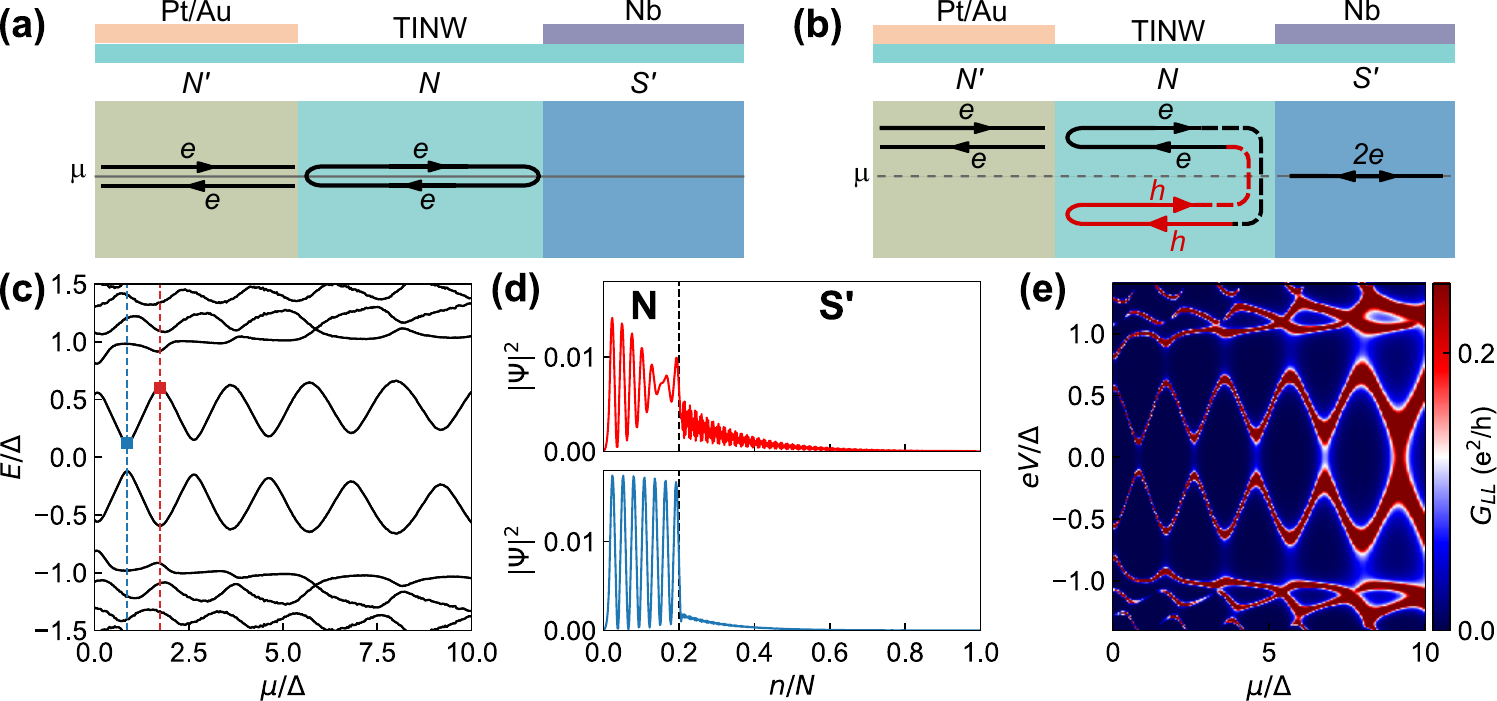}
	\caption{\linespread{1.05}\selectfont{}
	{\bf Reflection processes and simulated Andreev bound states in nanowires. }
	{\bf (a, b)} Schematic drawings of the normal reflection and Andreev reflection processes that contribute to the formation and energy of ABSs in the bare section (N) of the TINW. Electron transfer from Pt/Au and from Nb causes the TINW sections beneath them (N' and S', respectively) to have different chemical potentials, resulting in normal reflection at both interfaces to the N section as well as Andreev reflection at the NS' interface. {\bf (c)} Simulated energy of an individual ABS in the N region (see supplement for details); the energy of the low-lying ABS shown here is given by the relative ratio of Andreev reflection and normal reflection. Normal reflection is largest at particular resonant $k_F$ values satisfying $\cos (2k_F L) = 0$ due to Fabry-Perot-like interference in the quantum well. This results in a minimum of the ABS energy as a function of chemical potential $\mu$ at on resonance (blue dashed line), and a maximum at off resonance (red dashed line). {\bf (d)} The wavefunction of the ABS at off resonance (red) has a significant weight in the S' section of the nanowire (right of black dashed line). In contrast, when on resonance, the weight of the ABS in the S' region is significantly reduced. {\bf (e)} Due to the ABS in the junction region, the simulated local conductance $G_{\rm LL}$ exhibits oscillations as a function of chemical potential. Note that in the experimental setup there is no explicit tunnel barrier and so the broadening of the conductance is primarily dependent on the coupling to the normal lead (here largest for large $\mu$). Large broadening of the ABS energy results in a finite conductance at zero-bias.}
    \label{fig:Figure2}
\end{figure}

\begin{figure}[ht]
    \centering
    \includegraphics[width=0.85\linewidth]{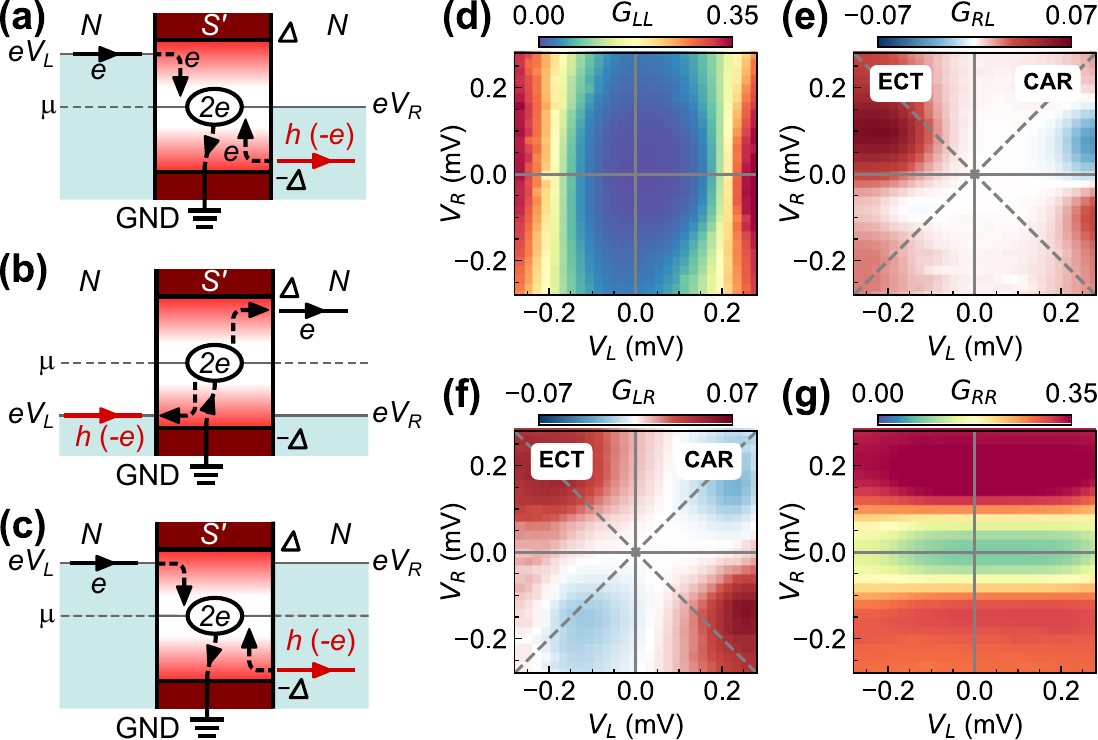}
	\caption{\linespread{1.05}\selectfont{}
        \textbf{Nonlocal processes involving Cooper pairs and conductance matrix as a function of $V_{\rm L}$ and $V_{\rm R}$. }
        {\bf (a)} Schematic drawing of the crossed Andreev reflection (CAR) process in S' contacted by N on both sides.
        \YA{Red colour indicates the presence of Andreev/bulk states within S' which itself is proximitized by Nb.}
         An electron with energy $eV_{\rm L}$ coming from the left N creates a Cooper pair in S' by taking an electron from the right N, emitting a hole with energy $-eV_{\rm L}$ to the right. The created Cooper pair is drained to the ground. {\bf (b)} Cooper-pair splitting: When S' in the middle is positively biased, the probability of splitting a Cooper pair (provided from the ground) into the two N contacts is maximal. {\bf (c)} Cooper-pair creation: When S' is negatively biased, the probability to create a Cooper pair in S' by taking electrons from the two N contacts is maximal, and the Cooper pair is drained to the ground. The Cooper-pair splitting/creation is microscopically a CAR process.
        \textbf{(d)} $G_{\rm LL}$, \textbf{(e)} $G_{\rm RL}$, \textbf{(f)} $G_{\rm LR}$, \textbf{(g)} $G_{\rm RR}$ measured at 17 mK by varying both $V_{\rm L}$ and $V_{\rm R}$ at a fixed $V_{\rm g} = -4.368$ V. Note the different colour scales for local and nonlocal conductances. CAR is enhanced when $V_L$ and $V_R$ have the same sign, while opposite signs between $V_L$ and $V_R$ promotes ECT \cite{Bordin2023}. 
        \YA{The behavior of $G_{\rm LR}$ in panal (f) indeed follow this expectation. The $G_{\rm RL}$ data in panel (e) slightly deviate from the expectation, which is likely due to the non-negligible $G_{\rm RR}$ at $V_R$ = 0 that would enhance ECT. The data of $G_{\rm LL}$ and $G_{\rm RR}$ in panels (d) and (g) confirm that the local conductances are not affected by the bias on the other side. 
   } }
    \label{fig:Fig3}
\end{figure}

\begin{figure}[ht]
    \centering
    \includegraphics[width=\linewidth]{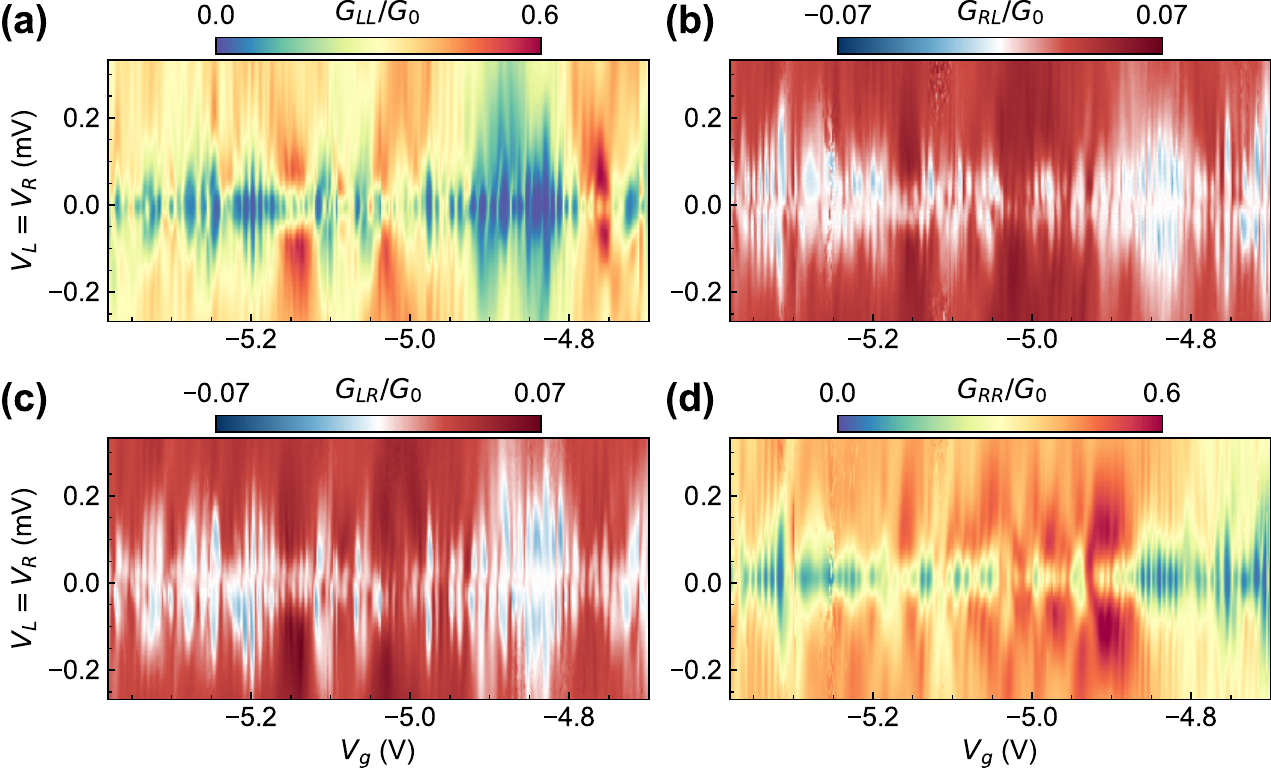}
	\caption{\linespread{1.05}\selectfont{}
        {\bf  Conductance matrix measured with $V_R=V_L$ in zero magnetic field.}
        {\bf (a)-(d)} $G_{\rm LL}$, $G_{\rm RL}$, $G_{\rm LR}$, and $G_{\rm RR}$ measured at 17 mK in 0 T as a function of $V_{\rm g}$ and $V_{\rm L}$ (= $V_{\rm R}$). The $V_{\rm g}$ range shown here is near the Dirac point.  For the measurement of $G_{\rm RL}$, $V_{\rm L}$ is modulated with a frequency $f_L$ and the ac component in $I_R$ with the frequency of $f_L$ is lock-in detected. This way, $G_{\rm RL}$ measured on the right reflectes the ac drive on the left. The same is true for $G_{\rm LR}$.  
    }
    \label{fig:Fig4}
\end{figure}

\begin{figure}[ht]
    \includegraphics[width=\linewidth]{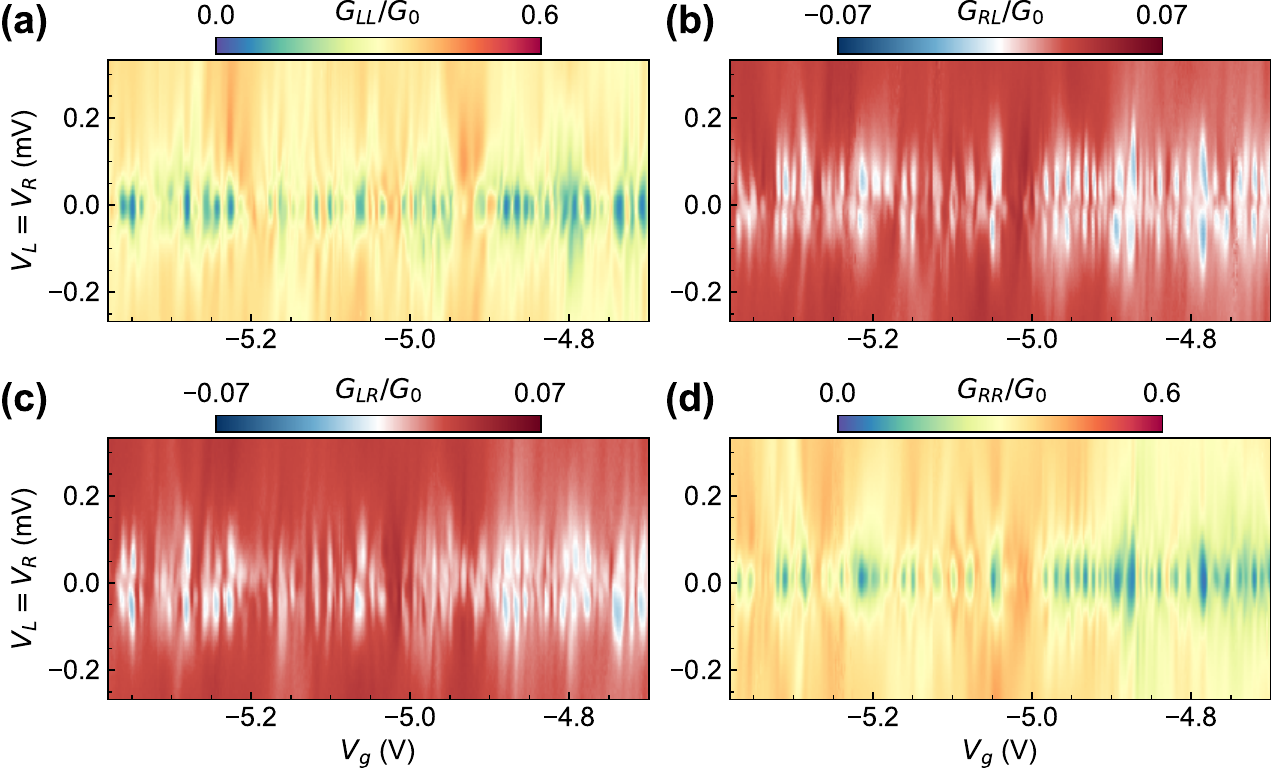}
	\caption{\linespread{1.05}\selectfont{}
        {\bf  Conductance matrix \YA{measured with $V_R=V_L$} for $\varphi = \frac{1}{2}(h/e)$.}
        \textbf{(a)-(d)} Similar set of data as in Fig. 4 but in the parallel magnetic field of 1.35 T which generates the magnetic flux $\varphi = \frac{1}{2}(h/e)$ in the TINW. The gaps in $G_{\rm LL}$ and $G_{\rm RR}$ due to the ABS, as well as the CAR signal in $G_{\rm LR}$ and $G_{\rm RL}$, are both observed at this flux, implying that the proximity-induced superconductivity is not destroyed by the angular-momentum mismatch. Nevertheless, the spectral features are weakened because of the softer SC gap.
    }
    \label{fig:Fig5}
\end{figure}

\begin{figure}[ht]
    \includegraphics[width=\linewidth]{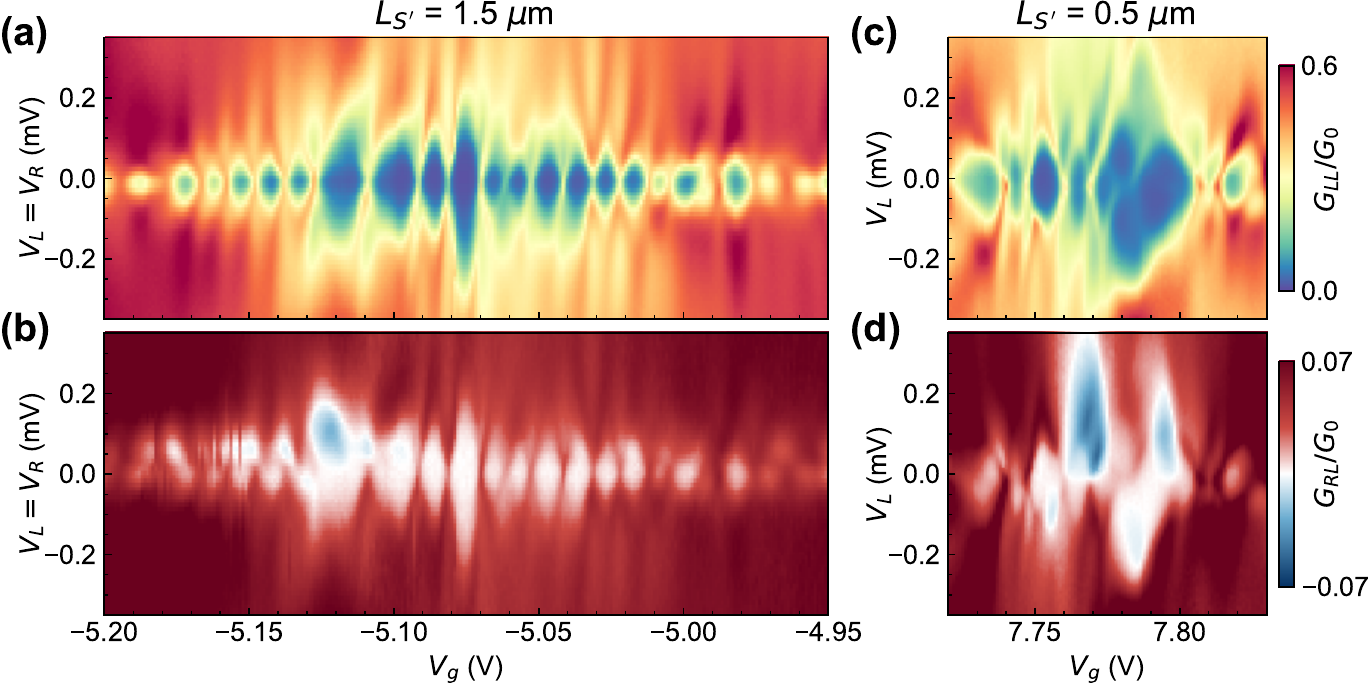}
	\caption{\linespread{1.05}\selectfont{}
        \textbf{$G_{LL}$ and $G_{RL}$ in devices having different lengths of the S'-section. }
        \textbf{(a, b)} $G_{LL}$ and $G_{RL}$ in a device with $L_{S'}$ = 1.5 $\mu$m measured with $V_R = V_L$ for a $V_{\rm g}$ range from $-5.20$ to $-4.95$ V. 
        \textbf{(c, d)} Similar data in a device with $L_{S'}$ = 0.5 $\mu$m for a $V_{\rm g}$ range from $7.72$ to $7.83$ V.  The sizes of the horizontal axis are tuned such that the same horizontal length in the plots corresponds to the same $V_{\rm g}$ range for the two data sets.  Note that the $V_{\rm g}$-distance between the resonances is essentially independent of $L_{S'}$. Negative $G_{RL}$ was observed even for $L_{S'}$ = 1.5 $\mu$m, whereas $\xi_{\rm SC}$ is estimated to be no longer than 110 nm. 
    }
    \label{fig:Fig6}
\end{figure}

\end{document}


\maketitle

\section{Measurement circuit}

\begin{figure}[ht]
    \centering
    \includegraphics[width=0.8\linewidth]{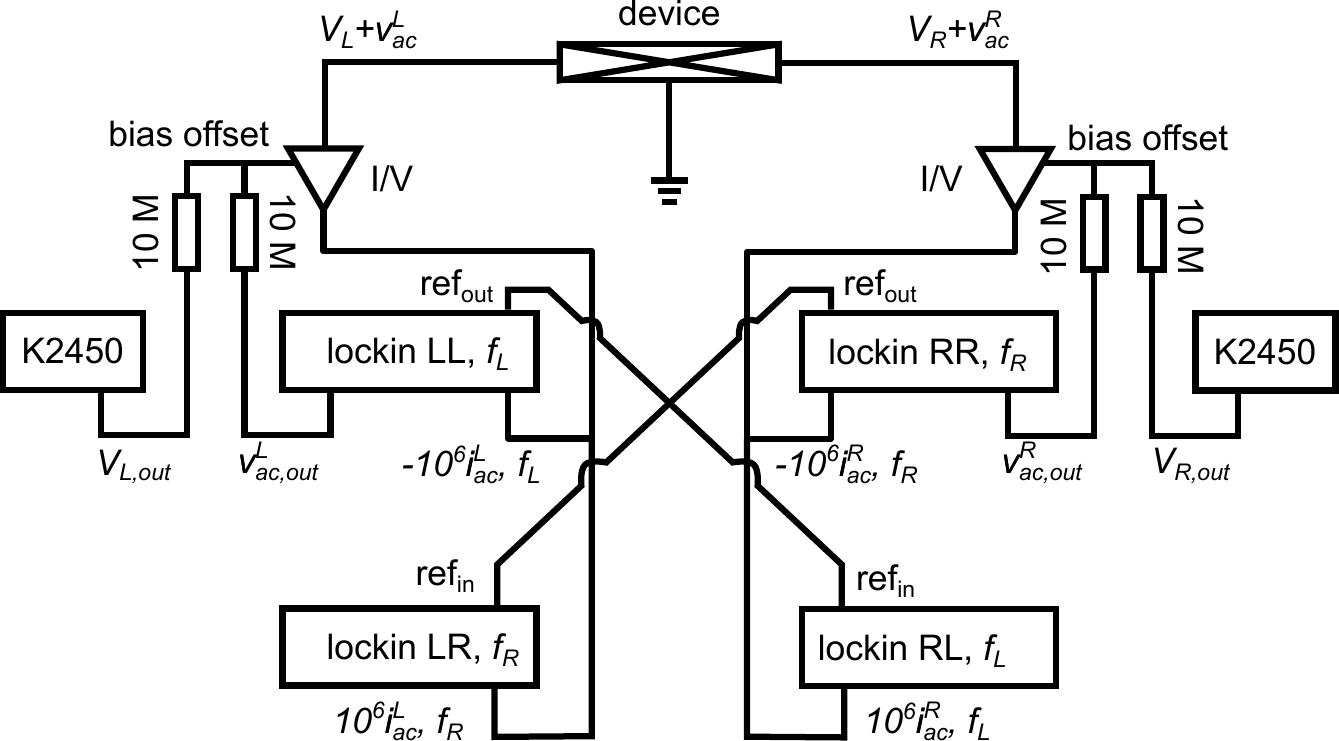}
    \caption{\linespread{1.05}\selectfont{}
        \textbf{Sketch of the measurement circuit. } 
        By using different frequencies for the left and the right electrodes, we can measure the conductance matrix in the same circuit\cite{Gramich2017, Menard2020}. We use Keithley DC source K2450 to generate DC voltages and lock-ins (Stanford Research Systems SR860 or NF LI5640) to generate ac voltages. The voltage outputs from these instruments, $V_{\rm L, out}$ of DC voltage and $v_{\rm ac, out}^{\rm L}$ of ac voltage at frequency $f_{\rm L}$, are applied to 10-M$\Omega$ resistors separately, and then supplied to the bias offset port of the amplifier LSK389A from Basel Precision Instruments. The bias offset port is grounded via a 10-k$\Omega$ resistor, such that $V_{\rm L, out}$ and $v_{\rm ac, out}^{\rm L}$ are converted to $V_{\rm L} = V_{\rm L, out}/1000$ and $v_{\rm ac}^{\rm L} = v_{\rm ac, out}^{\rm L}/1000$ and the total bias voltage $V_{\rm L}+ v_{\rm ac}^{\rm L}$ is applied to the sample through the amplifier, which is connected to the left electrode of the device under test. The local ac current $i_{\rm ac}^{\rm L}$ is amplified by the left amplifier and measured by the lock-in named ``LL'' at the frequency $f_{\rm L}$. The corresponding nonlocal ac current $i_{\rm ac}^{\rm R}$, which is driven by $V_{\rm L}+ v_{\rm ac}^{\rm L}$ at the frequency $f_{\rm L}$, is amplified by the right amplifier and measured by the lock-in named ``RL'' using the reference signal taken from the lock-in ``LL''. The $G_{\rm LL}$ and $G_{\rm RL}$ can be calculated with $G_{\rm LL}=i_{\rm ac}^{\rm L}/v_{\rm ac}^{\rm L}$ and $G_{\rm RL}=i_{\rm ac}^{\rm R}/v_{\rm ac}^{\rm L}$, both measured at the frequency $f_{\rm L}$. 
The $G_{\rm RR}$ and $G_{\rm LR}$ are measured and calculated in a symmetric manner but with a different ac frequency $f_{\rm R}$. 
    }
    \label{fig:Supp1}
\end{figure}

\newpage
\section{Background subtraction in the AB-oscillation data}

\begin{figure}[ht]
    \centering
    \includegraphics[width=0.95\linewidth]{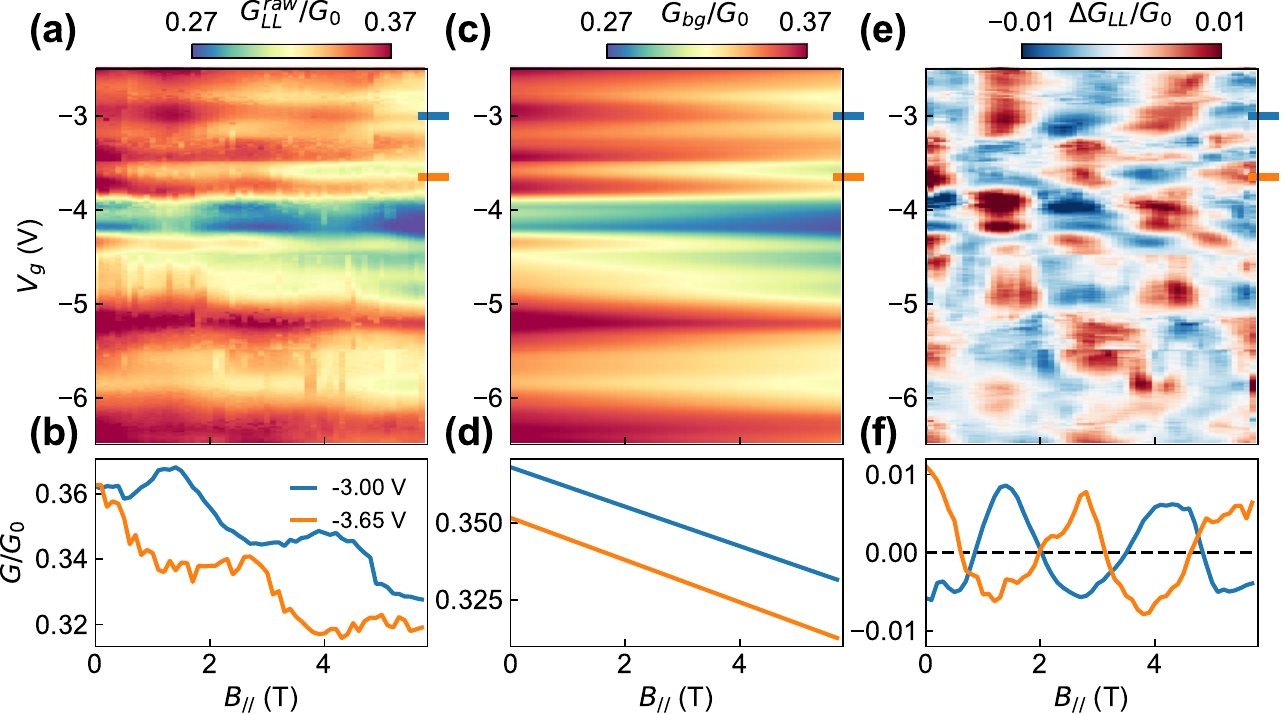}
    \caption{\linespread{1.05}\selectfont{}
        \textbf{(a,b)} Raw data of the ``checker-board'' pattern shown in Fig. 1b of the main text. The $V_{\rm g}$-dependent oscillations are already visible in the raw data. Two representative line-cuts at $V_{\rm g}$ = $-3.00$ V and $-3.65$ V are plotted in (b), where one can see the Aharonov-Bohnm (AB)-like oscillations with phase shifts. 
        \textbf{(c,d)} We fitted a linear $B_{\parallel}$-dependence $G_{\rm bg} = aB_{\parallel} + b$ for the background $G_{\rm bg}(B_{\parallel})$ at each $V_{\rm g}$ and subtracted it from the raw $G_{\rm LL}^{\rm raw}(B_{\parallel})$ data to obtain $\Delta G_{\rm LL} \equiv G_{\rm LL}^{\rm raw} - G_{\rm bg}$ at each $V_{\rm g}$.  The $G_{\rm bg}(B_{\parallel}, V_{\rm g})$ behavior used for the analysis is shown in (c) and the line-cuts at $V_{\rm g}$ = $-3.00$ V and $-3.65$ V are plotted in (d). 
        \textbf{(e,f)} The obtained $\Delta G_{\rm LL}$ and two line-cuts at $V_{\rm g}$ = $-3.00$ V and $-3.65$ V. Noise spikes caused by the gate sweeping have been removed from the $\Delta G_{\rm LL}$ data using the Savitzky–Golay filtering (window length 7 data points, order 3). 
    }
\end{figure}

\newpage
\section{Temperature dependence of the normal-state oscillations}

\begin{figure}[h]
    \centering
    \includegraphics[width=0.84\linewidth]{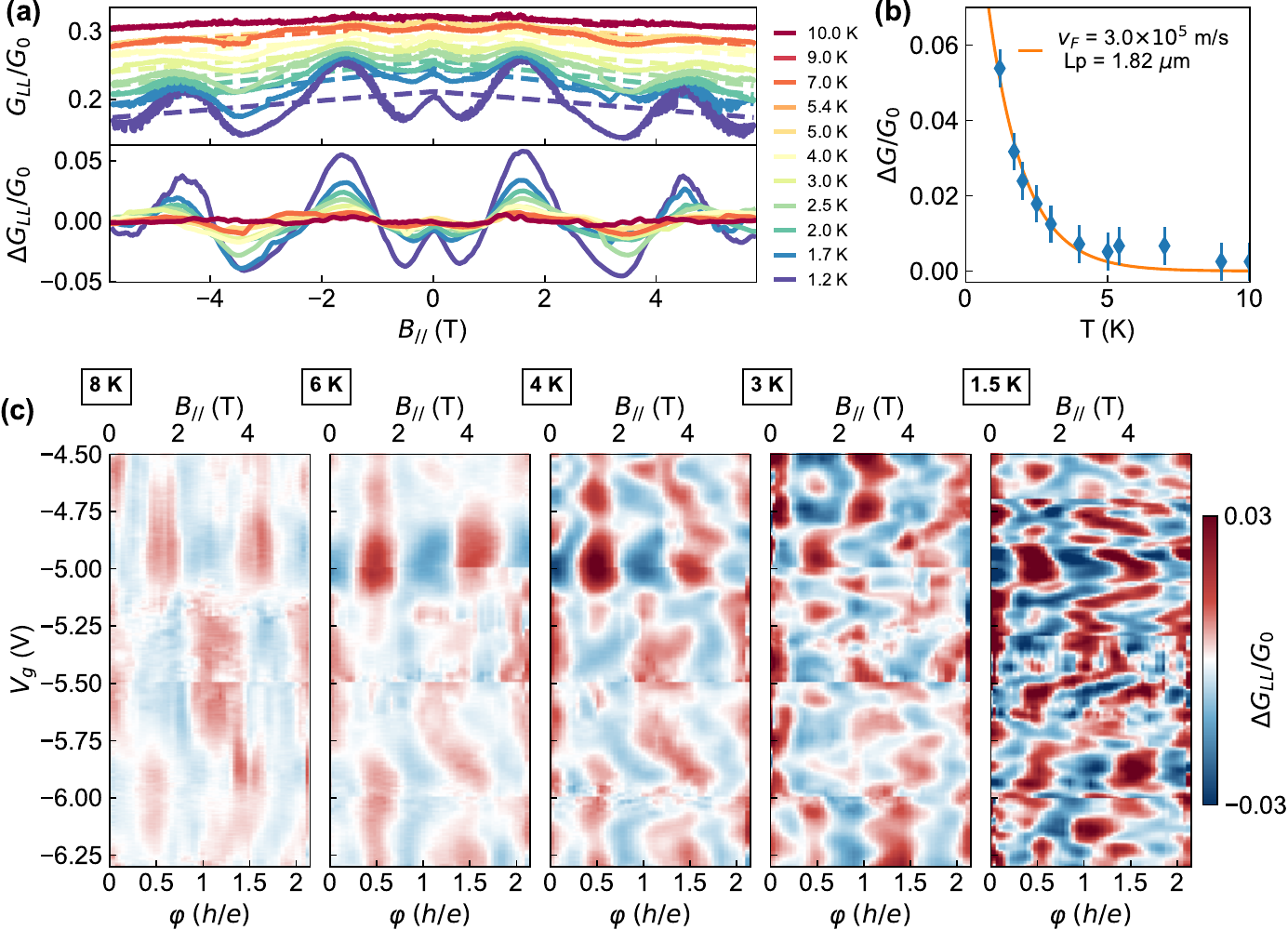}
    \caption{\linespread{1.05}\selectfont{}
        {\bf (a)} AB-like oscillations at $V_{\rm g}$ = $-4.37$ V. In the upper panel, the raw data (solid lines) and the linear background (dashed lines) are shown for various temperatures from 1.2 to 10 K. In the lower panel, the $\Delta G_{\rm LL}$ data after the background subtraction \YA{and Savitzky-Golay filtering} are plotted. 
        {\bf (b)} Oscillation amplitude extracted at $\varphi=-1.6$ T, where the amplitude was maximum. 
        \YA{The error bars are due to the error in the magnitude of $i_{\rm ac}$ caused by the noise in the lock-in measurements, and they are estimated to be the same $0.005 G_0$ for all the data points.}
      By using the formula\cite{Dufouleur2013} $\Delta G_{\rm LL}\varpropto \exp{(-k_B TL_{\rm p}/\hbar v_{\rm F})}$, we can extract the phase coherent length $L_{\rm p} = 1.82$ $\mu$m for $v_{\rm F}$ = 3$\times$10$^5$ m/s. 
        {\bf (c)} Change of the checker-board pattern upon decreasing the temperature from 8 K to 1.5 K. The checker-board pattern breaks up into more fine-structured patterns at lower temperature, which is presumably due to the resonance states caused by Fabry-Perot-like interference in the N-section of the TINW, which have the energy-level spacing of about 0.5 meV (see main text).}
    \label{fig:Supp5}
\end{figure}

\newpage
\section{Conductance matrix in parallel magnetic fields}

\begin{figure}[ht]
    \centering
    \includegraphics[width=0.95\textwidth]{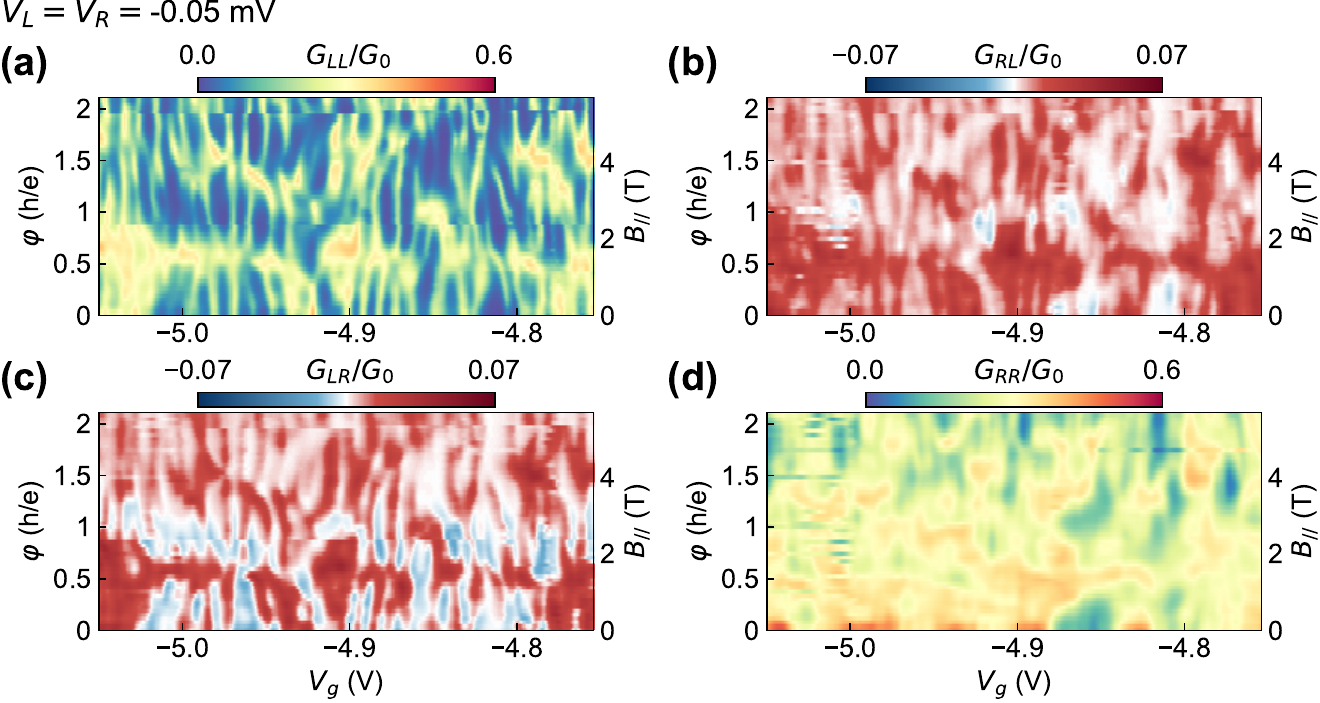}
    \caption{\linespread{1.05}\selectfont{}
        \textbf{(a)-(d)} $G_{\rm LL}$, $G_{\rm RL}$, $G_{\rm LR}$ and $G_{\rm RR}$ measured at 17 mK with $V_{\rm L} = V_{\rm R} = -0.05$ mV as a function of $\varphi$ (magnetic flux created by $B_{\parallel}$) and $V_{\rm g}$. The small bias voltage is applied to drive the device into the Cooper-pair splitting regime, such that the CAR process in $G_{\rm RL}$ and $G_{\rm LR}$ is promoted, while keeping $G_{\rm LL}$ and $G_{\rm RR}$ to be similar to their zero-bias values. At around $\varphi=\frac{1}{2}(h/e)$, $G_{\rm RL}$ and $G_{\rm LR}$ are mostly dominated by ECT, but CAR is still visible at some $V_{\rm g}$ values, suggesting that the induced superconductivity in S' does not disappear for $\varphi=\frac{1}{2}(h/e)$. 
    }
\end{figure}

\newpage
\section{$G_{\rm RR}$ and $G_{\rm LR}$ in the 1.5 $\mu$m device}
\vspace{-5mm}

\begin{figure}[ht]
    \centering
    \includegraphics[width=0.75\linewidth]{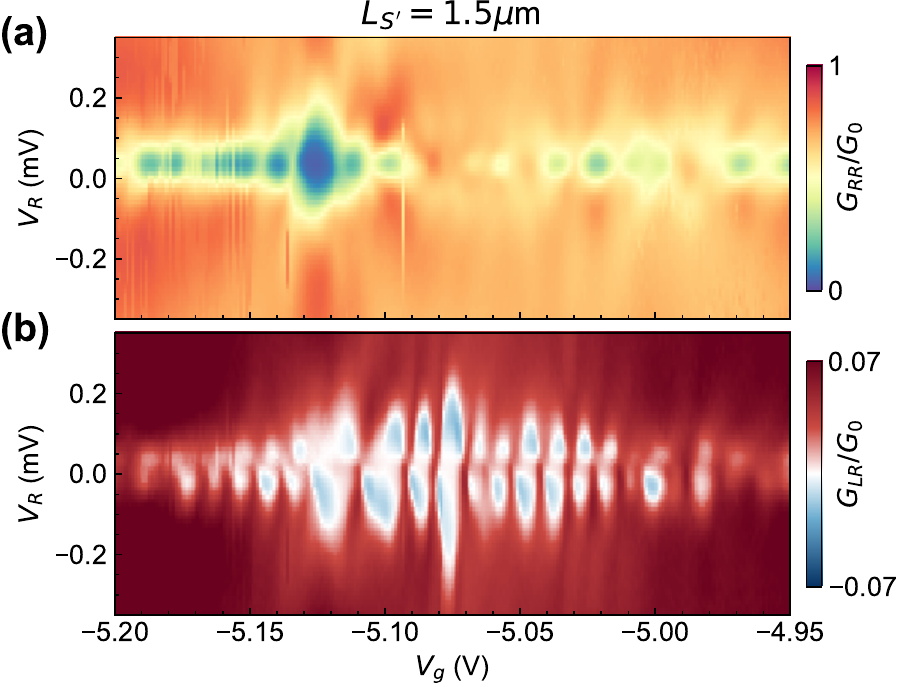}
    \vspace{-3mm}
    \caption{\linespread{1.05}\selectfont{}
        \textbf{(a)} $G_{\rm RR}$ and \textbf{(b)} $G_{\rm LR}$ of the 1.5 $\mu$m device measured at 17 mK in 0 T with $V_{\rm R} = V_{\rm L}$.  
    }
    \label{fig:FigS_15um_GRR}
\end{figure}

In Fig. 6 of the main text, the $G_{\rm LL}$ and $G_{\rm RL}$ data obtained from the 1.5 $\mu$m device are shown. To present the complete conductance matrix, Fig. \ref{fig:FigS_15um_GRR} shows the remaining two components, $G_{\rm RR}$ and $G_{\rm LR}$, measured at the same time. Negative $G_{\rm RL}$ signifying the CAR process is clearly observed.

\vspace{5mm}

\newpage
\section{Modeling of chemical-potential-dependent local conductance}

\begin{figure}[bh]
    \centering
    \includegraphics[width=1\linewidth]{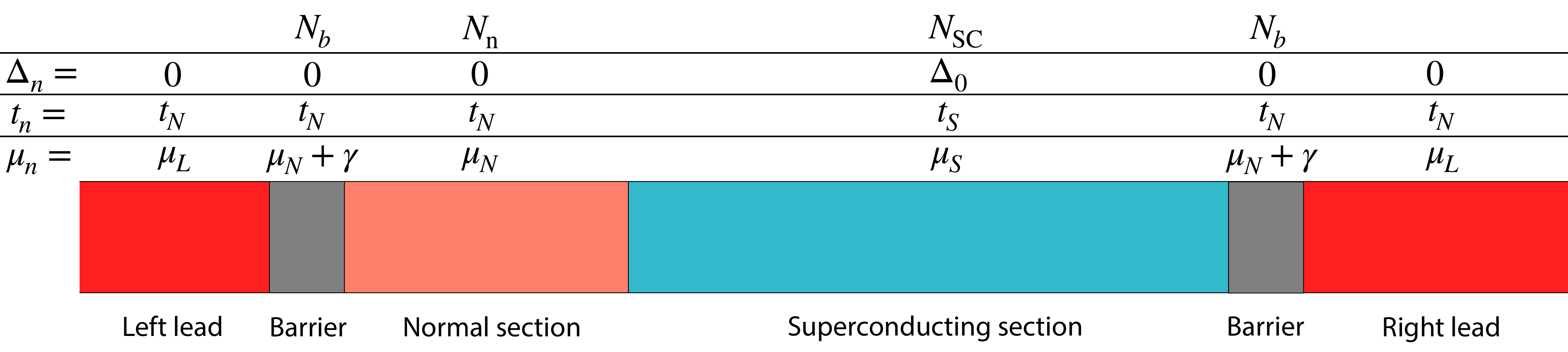}
    \caption{\linespread{1.05}\selectfont{}Schematic of one-dimensional nanowire model hosting an Andreev bound state in normal section. The number of sites and the correspondence of parameters $t_n$, $\Delta_n$, and $\mu_n$ in Eq.~\eqref{Ham} are given in the table above the relevant section.}
    \label{fig:FigS_nanowire_model}
\end{figure}

\vspace{-3mm}
To model the Fabry-Perot-like oscillations of the Andreev bound state (ABS) energy and corresponding local conductance, $G_{LL}$, shown in Fig. 2 of the main text, we use the software package KWANT\cite{KWANT}. In particular, we employ a one-dimensional model similar to that discussed in Refs.~\citenum{Reeg2018}~\&~\citenum{Hess2021}, such that the Hamiltonian of the one-dimensional nanowire is given by
\begin{equation}
\begin{aligned}
H= & \sum_{n=1}^N\left[\sum _ { \sigma}  \left( c _ { n , \sigma } ^ { \dagger } \left\{t_{n+\frac{1}{2}}+t_{n-\frac{1}{2}}-\mu_n+\gamma_n\right\} c_{n, \sigma} - t_{n+\frac{1}{2}} c_{n, \sigma}^{\dagger} c_{n+1, \sigma}\right) \right. \left.+\Delta_n c_{n, \downarrow}^{\dagger} c_{n, \uparrow}^{\dagger}+\text { H.c. }\right],
\end{aligned}\label{Ham}
\end{equation}
here $c_{n, \sigma}^{\dagger}$ ($c_{n, \sigma}$) creates (annihilates) an electron with spin $\sigma=\uparrow, \downarrow$ at the lattice site $n$. The total number of sites is given by $N=N_n+2N_b+N_{\rm SC}$, where $N_n$ is the number of sites in the normal section on the left of the nanowire, $N_b$ is the number of sites in the tunnel barriers on the left and right of the nanowire that are used to tune the strength of coupling to the leads, and $N_{\rm SC}$ is the number of sites in the superconducting section (see Fig.~\ref{fig:FigS_nanowire_model}). For our simulations we choose $(N_{\rm SC},N_n,N_b)=(400,98,9)$ sites. Although our aim is to provide a minimal model of  Fabry-Perot-like oscillations and not simulate our precise experimental setup, for our chosen parameters these number of sites will broadly correspond to the lengths of the corresponding sections in our experiment (see below). 
\YA{Note that the quantization of the transverse modes into subbands indexed by angular momentum assures the separability of the transverse physics from longitudinal physics in TINW\cite{Legg2021, Legg2022}, justifying the use of the one-dimensional model to capture the key features in the longitudinal physics.}

The parameters $t_n$ and $\mu_n$ in Eq.~\eqref{Ham} denote the nearest-neighbour hopping matrix element and the chemical potential, respectively. The value of these depends on the specific section of the nanowire. In particular, when $n$ corresponds to a normal site or the tunnel barrier the hopping is given by $t_n=t_N$ and in the superconducting section it is given by $t_n=t_S$. Similarly, the chemical potential in the normal section is given by $\mu_n=\mu_N$, in the superconducting section by $\mu_n=\mu_S$, and in the tunnel barriers by $\mu_n=\mu_N+\gamma$, with $\gamma$ the barrier height. In our experiment there is no well defined tunnel barrier so in our simulations we consider only small barrier heights $\gamma$ with large broadening by the leads (see below) . The superconducting pairing potential is $\Delta_n=\Delta_0$ in the superconducting section and is zero otherwise. Note, for simplicity we do not include a spin-splitting of the band due to spin-orbit coupling (SOC) since it is unimportant for our discussion of the qualitative features due to Fabry-Perot physics. 
\YA{The inclusion of SOC does not alter the main mechanism of the Fabry-Perot resonance\cite{Hess2021}, but SOC is important for the realization of the topological phase\cite{Reeg2018} and its spatial variation is also important for the emergence of the Andreev bands\cite{Hess2021} in Rashba nanowires. In TINWs, the way how SOC enters the problem is different but its role in the topological physics is similar\cite{Legg2021, Legg2022}.}

For the simulations in the main text, we choose $(t_N,t_S)=(50,25)$ meV, $\Delta_0=0.4$ meV, $(\mu_N,\mu_S,\gamma)=(2+\mu,40,5)$ meV where $\mu$ adjusts the chemical potential in the normal section and barrier (see Fig. 2 of main text). As above, we note that, although our focus in these simulations is on the qualitative features, our chosen values are broadly consistent with the length-scales and energies expected in our experiment. For instance, in our experiment the superconducting section has a length $L_S\sim 1.5$ $\mu$m which would correspond to a lattice spacing of $a=L_S/N_{\rm SC} =3.5$ nm resulting in a Fermi-velocity in the superconducting section of $v_F=2\sqrt{t_{S} \mu_{S}} a /\hbar \approx 3.5 \times 10^5$ m/s.

To compute conductance normal leads are attached on the left and right ends of the nanowire and are modelled by the same Hamiltonian as the normal sections ($\Delta_n=0$, $t_n=t_N$) but with fixed chemical potentials $\mu_L$. For our simulations we take $\mu_L=40$ meV i.e. the same as in the superconducting section. Throughout we focus on the local conductance in the left lead only since this is where we model the normal section at the junction to the lead, an equivalent normal section on the right of the nanowire would yield identical results for $G_{RR}$. As discussed in the main text, attaching leads broadens the ABS energy which, depending on the strength of the coupling of the ABS to the lead, can result in local conductance at zero-bias, even if the ABS energy does not reach zero energy when no leads are attached. Although the majority of broadening is due to strong coupling to the leads, further a small broadening also occurs due to temperature, which we include via the Fermi-Dirac distribution\cite{Hess2021} and take to be $T=20$ mK.

Finally we note that we are primarily interested in the qualitative physics of Fabry-Perot-like oscillations of ABS energy and, as such, here we utilise a purely one-dimensional model for a nanowire with a single (spin degenerate) subband that results in a single (spin degenerate) Andreev bound state in the normal section. This is the minimal model required to reproduce oscillations in local conductance as a function of chemical potential that are observed in our experiment. In reality, however, there will be multiple subbands each with different $v_F$ and $k_F$ values. Each subband will contribute to the Andreev bound state spectrum, with all contributions exhibiting Fabry-Perot-like oscillations of energy with various amplitudes, chemical potential dependences, and broadening from the leads. The simulation results are shown in Figs. 2c-2e of the main text.

\newpage
\section{\YA{Zoom-in on the data shown in Fig. 1c of the main text}}

\begin{figure}[ht]
    \centering
     \YA{\includegraphics[width=0.9\textwidth]{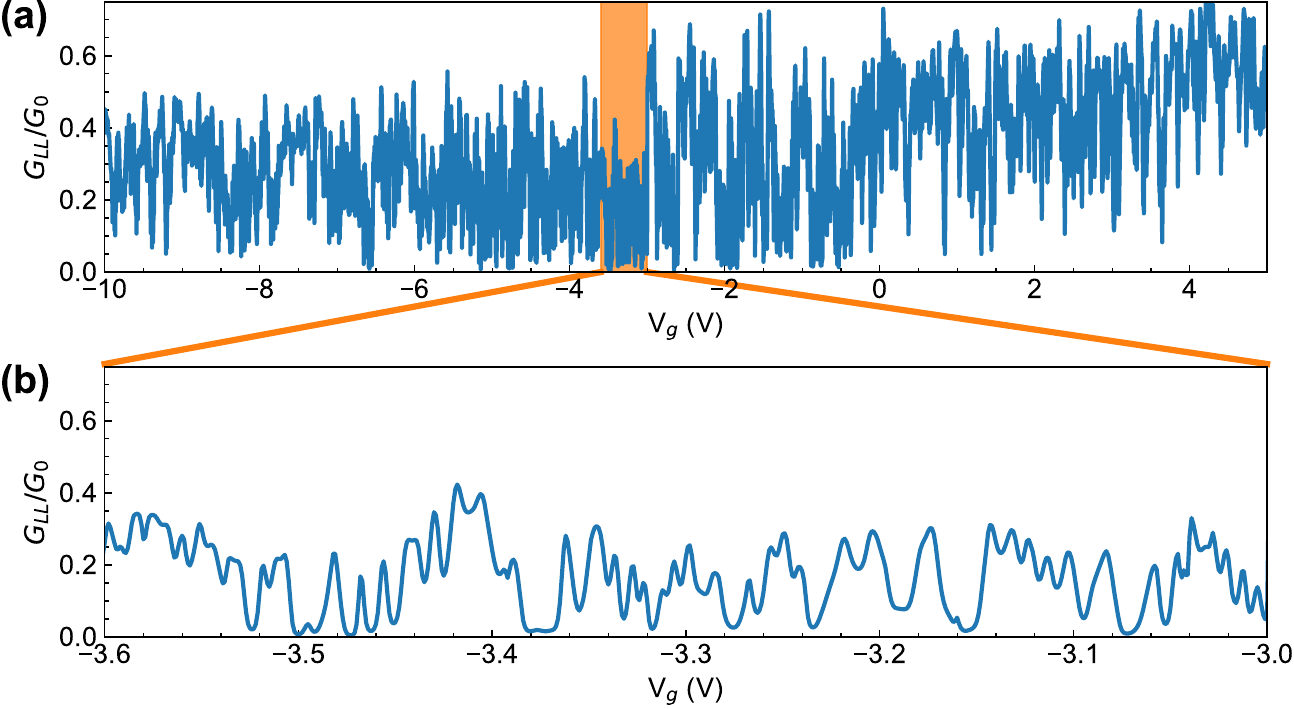}
    \caption{\linespread{1.05}\selectfont{}
      Zoom-in of the $G_{\rm LL}$ vs $V_{\rm g}$ data presented in Fig. 1c of the main text for a narrower $V_{\rm g}$-range of $-3.6$ V to $-3.0$ V.
    }
    \label{fig:FigS_zoomin}
    }
\end{figure}

\YA{Since the 17-mK data shown in Fig. 1c of the main text look like noise, to demonstrate that the strong conductance fluctuations are not due to noise but due to fast $V_{\rm g}$-dependent oscillations, a zoom-in on the data is presented in Fig.~\ref{fig:FigS_zoomin}. Note that the measurements for Fig. 1c and Fig. 1d of the main text were performed separately, and the unavoidable gate-sweep hysteresis makes the details of the two data set slightly different.}

\newpage
\section{\YA{Asymmetry between $G_{\rm LR}$ and $G_{\rm RL}$ in Figs. 3e and 3f of the main text}}

\YA{It is known that differences in the local conductance can affect the nonlocal conductance, such that $G_{\rm LR}$ and $G_{\rm RL}$ are not the same (but must still respect certain relations from symmetries). Detailed discussions on the conductance symmetries of multiterminal devices can be found in Refs. \citenum{Danon2020, Maiani2022}. In the present case, we found experimentally that $G_{\rm RL}$ and $G_{\rm LR}$ depend strongly on the local conductance where the current is measured. For example, for $G_{\rm LR}$ shown in Fig. 3f of the main text, the current was measured from the left, where $G_{\rm LL}$ was off-resonance and the ECT rate was low, making CAR to be easily visible. On the other hand, for $G_{\rm RL}$ shown in Fig. 3e, the current was measured from the right, where $G_{RR}$ was not exactly at off-resonance and the ECT rate was higher; this caused a positive offset in $G_{\rm RL}$, smearing the CAR signal.}